\documentclass{vgtc}                          

\ifpdf
  \pdfoutput=1\relax                   
  \usepackage{graphicx}                
  \DeclareGraphicsExtensions{.pdf,.png,.jpg,.jpeg} 
\else
  \usepackage{graphicx}                
  \DeclareGraphicsExtensions{.eps}     
\fi%

\graphicspath{{figures/}{pictures/}{images/}{./}} 

\usepackage{microtype}                 
\PassOptionsToPackage{warn}{textcomp}  
\usepackage{textcomp}                  
\usepackage{mathptmx}                  
\usepackage{times}                     
\usepackage{cite}                      
\usepackage{tabu}                      
\usepackage{booktabs}                  
\usepackage{multirow}%

\onlineid{0}

\vgtccategory{Research}

\vgtcinsertpkg

\title{Quantifying the Impact of XR Visual Guidance on User Performance
\\ Using a Large-Scale Virtual Assembly Experiment}

\author{Leon Pietschmann
\thanks{e-mail: LHP30@cantab.ac.uk}\\ %
    \parbox{1.4in}{\scriptsize \centering University of Cambridge \\ and Harvard University}
\and Paul Zürcher\\ %
    \parbox{1.4in}{\scriptsize \centering University of Cambridge}
\and Erik Bubik\\ %
    \parbox{1.4in}{\scriptsize \centering University of Cambridge}
\and Zhutian Chen\\ %
    \parbox{1.4in}{\scriptsize \centering Harvard University}
\and Hanspeter Pfister\\ %
    \parbox{1.4in}{\scriptsize \centering Harvard University}
\and Thomas Bohné\\ %
    \parbox{1.4in}{\scriptsize \centering University of Cambridge}
    }

\teaser{
    \includegraphics[width=0.60\textwidth]{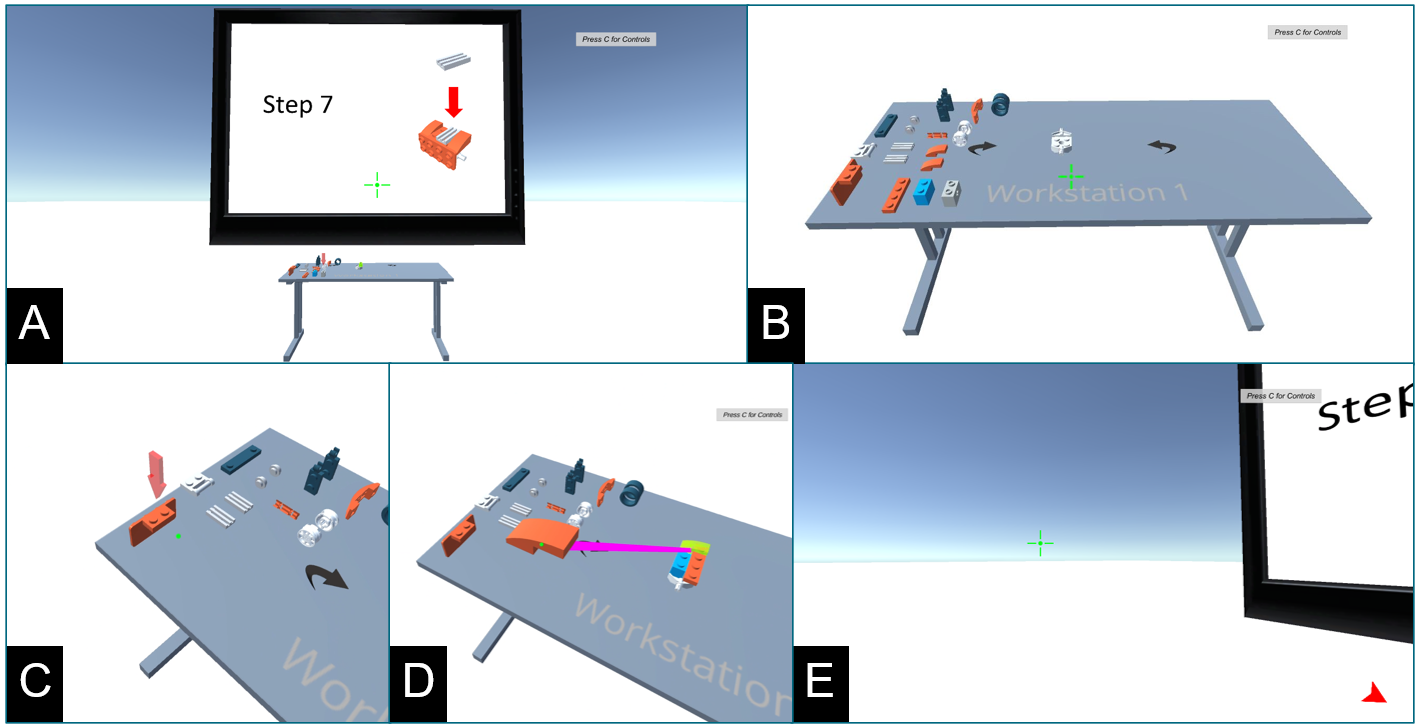}
    \label{fig:1}
    \centering
    \caption{Overview of the virtual environment, featuring the assembly instruction board (A), one of three workstations (B), a red guidance arrow to support Object Identification (C), a purple connection line and green placement hologram to support Action Guidance (D), and a red indicator at the edge of the screen to support Gaze Guidance (E).}
}

\abstract{The combination of Visual Guidance and Extended Reality (XR) technology holds the potential to greatly improve the performance of human workforces in numerous areas, particularly industrial environments. Focusing on virtual assembly tasks and making use of different forms of supportive visualisations, this study investigates the potential of XR Visual Guidance. Set in a web-based immersive environment, our results draw from a heterogeneous pool of 199 participants. This research is designed to significantly differ from previous exploratory studies, which yielded conflicting results on user performance and associated human factors. Our results clearly show the advantages of XR Visual Guidance based on an over 50\% reduction in task completion times and mistakes made; this may further be enhanced and refined using specific frameworks and other forms of visualisations/Visual Guidance. Discussing the role of other factors, such as cognitive load, motivation, and usability, this paper also seeks to provide concrete avenues for future research and practical takeaways for practitioners.} 

\CCScatlist{ 
  \CCScat{}{Human-centered computing}{Visualization, Human computer interaction (HCI), Interaction design}{Empirical studies in visualization, HCI, and interaction design}
  }

\begin{document}



\maketitle


\section{Introduction}

Extended Reality (XR) technologies such as Augmented or Virtual Reality (AR/VR) provide immersive and interactive ways to seamlessly experience a blend of real and virtual elements. They hence offer the potential to significantly impact a range of fields, including manufacturing, education, and healthcare~\cite{Jasche2021, Radu2019, Nilsson2007}. 
By embedding visual support within the user’s field of vision XR technologies may be leveraged to increase worker performance during both training and operations, e.g. by reducing the time required to complete tasks or the number of mistakes made in the process~\cite{Jasche2021}. 
Although the literature on XR technology is extensive, the respective influence, design, and implementation of Visual Guidance are not well-understood overall. Further, only few studies focus on comparing the impact of different visualisation types on user performance \textit{ceteris paribus}~\cite{Jasche2021}. 

Therefore, in this study, we quantify the impact of XR Visual Guidance on user performance using a large-scale virtual assembly experiment. 
To recruit a large number of participants, our research design was based on webVR, meaning an immersive but desktop-based setup. Given our focus on different forms of visualisations, this approach allowed for comparable insights while accessing a broader population of participants compared to more traditional lab-based XR Visual Guidance experiments; these are usually smaller in participant numbers and less diverse. To the best of our knowledge, 
this study is the first to quantify the effect of different forms of XR Visual Guidance on user performance on such a large scale.
 
In pursuing the following objectives, we hope to advance the fields of XR Visual Guidance, operator performance-oriented visualisation research, and Human-Computer Interaction by (1) quantifying the impact different modes of XR Visual Guidance have on user performance, (2) exploring the human factors tied to potential changes in performance, and (3) further refining and field-testing the XR Visual Guidance framework proposed by Pietschmann et al.~\cite{Pietschmann2022} as part of (4) a large-scale experiment.

\section{Background and Research Problem}

XR Visual Guidance refers to in-situ visual support in the form of information embedded into users’ field of vision via XR technology, e.g. AR headsets~\cite{Pietschmann2022, Jasche2021, tong_2022_exploring, chen_2022_augmenting, lin_2023_the, lin_2023_labeling}.
Previous studies have investigated the potential for applying XR Visual Guidance to a range of procedural knowledge-intensive tasks, including assembly, maintenance, and training~\cite{Pietschmann2022, Jasche2021, Blattgerste2018, Smith2020, Buchner2021}. While a multitude of studies have covered the effects of XR technology on user performance, much fewer have delved into the impact of XR Visual Guidance. Despite often reporting conflicting results on the relationship between XR Visual Guidance, user performance, and underlying human factors, the overall stance on XR Visual Guidance is positive~\cite{Pietschmann2022, Buchner2021, Akayr2017, Parong2021, chu_2022_tivee}.

It is hypothesised that well-applied XR Visual Guidance may increase overall user performance and benefit learning and retention~\cite{Pietschmann2022, Jasche2021, Bohne2022}.
This positive effect of Visual Guidance is often associated with a reduction in cognitive load~\cite{Blattgerste2017}.
Novelty effects, however, which stem from participants' unfamiliarity with XR hardware have been reported to negatively affect cognitive load. This causes opposing influences and facilitates conflicting results~\cite{Pietschmann2022}. To mitigate such novelty effects on cognitive load, previous studies recommended the implementation of pre-training into the research design. Doing so enables the familiarisation of participants with the virtual environment (VE), associated XR hardware, and the HCI process prior to the main study~\cite{Meyer2019}.
When it comes to the underlying task to be completed by the user, sequential tasks based on procedural knowledge have been among the most suitable applications of Visual Guidance~\cite{Pietschmann2022}.

Systematic approaches to XR Visual Guidance, such as the framework proposed by Pietschmann et al.~\cite{Pietschmann2022}, are still rare. 
Additionally, almost all related or similar empirical studies conducted small-scale (i.e. n $\approx$ 20) experiments, 
often based on highly specific tasks and with a largely male-dominated, undergraduate participant pool recruited from the respective researchers' university environments (for an in-depth discussion, see \cite{Pietschmann2022, Jasche2021}). 
The resulting lack of generalisability is driven by the exploratory experimental designs employed in previous studies in combination with largely homogeneous participant pools, ultimately making it challenging to quantify the impact XR Visual Guidance has on user performance~\cite{Pietschmann2022, Jasche2021}.
In this study, and based on related work, 
we investigate three hypotheses: We expect time to completion (\textbf{H1}), the number of mistakes (\textbf{H2}), and cognitive load (\textbf{H3}) all to significantly differ between the different visualisation types (see \autoref{fig:2}). \\

\section{Methodology}

In line with the XR Visual Guidance framework proposed by Pietschmann et. al~\cite{Pietschmann2022},
we investigated the impact of three different dimensions of Visual Guidance (VG): Gaze Guidance (GG), Object Identification (OI) and Action Guidance (AG) in a between-subjects design. 
The framework suggests that a sequential assembly task can be split into three repeating parts, 
which may be supported by distinct Visual Guidance: 
(1) orientation in the three-dimensional VE to move the field of vision in the correct general direction, supported by GG, 
(2) identification of the correct part to pick up, supported by OI, 
and (3) placing the part in the correct spot, supported by AG. 
For GG, we included a red arrow at the edge of the participant's field of vision, indicating the direction to look at (E in Fig. 1). 
OI was implemented as a red arrow pointing at the next part to pick up (C in Fig. 1). 
AG consisted of a transparent green hologram indicating the target location of where to add the part to the workpiece and a purple connection line between the target location and the part in hand, 
thus guiding the user to place the part (D in Fig. 1). 

\subsection{Study Design}
\begin{figure}[h]
    \centering
    \includegraphics[width=0.45\textwidth]{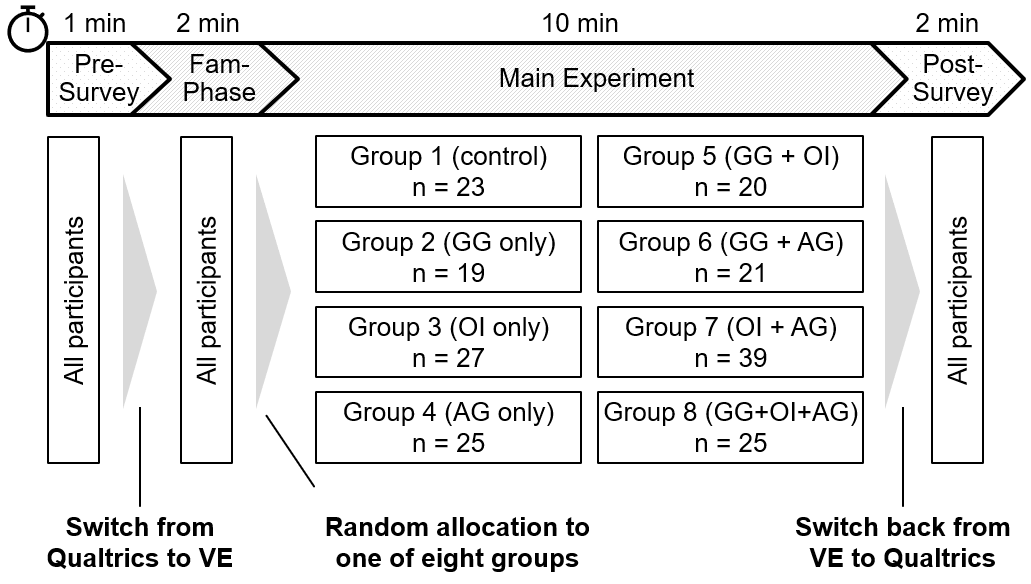}
    \caption{Experimental procedure and different groups including the respective number of participants (n) included in final analysis.}
    \label{fig:2}
\end{figure}

To measure the effect of every possible combination of GG, OI, and AG on the performance metrics, eight groups (G1-G8) were required, as shown in \autoref{fig:2}. An ex-ante power analysis yielded a total required number of participants greater than 168; hence, we opted for a web-based immersive experiment that could accommodate large amounts of participants, especially in times of COVID-19. We developed a system which automatically recruited participants via Amazon Mechanical Turk (MTurk), fed them through the experiment (including pre-and post-questionnaire and a browser-based virtual environment), and collected the data~\cite{zuercher2022optimising}.

Online XR experiments such as browser-based webVR applications have been increasingly recognised as a viable alternative to traditional in-lab XR research~\cite{Ratcliffe2021, tong2023understanding}. 
Although webVR simulations may not fully replicate those experiments using XR hardware,
they enable data collection from a greater and potentially more diverse pool of participants in less time, using fewer resources, 
while yielding comparable results – all of these factors were important for this study~\cite{Farr2023}.

Participants were recruited internationally and pseudonymously via MTurk and were allowed to participate only once based on their MTurk ID, which was their sole identifier.

\subsection{Procedure}
The experiment itself consisted of a pre-questionnaire, a familiarisation phase, one main assembly task, and a post-questionnaire. 
\autoref{fig:2} illustrates the experimental procedure. 
Firstly, after accepting the task, participants were forwarded to the pre-questionnaire hosted on Qualtrics. 
Following this, they received a link to the virtual environment which ran a full-screen webVR application on their machines. 
Secondly, participants were familiarised with the controls, interaction methods, an abstract assembly procedure, and specifics about the VE as well as the general goals of the experiment (Fam-Phase). We incorporated this pre-training principle to mitigate potential novelty effects on our results~\cite{Pietschmann2022}.
Thirdly, participants entered the main VE after successfully completing the familiarisation phase. They were randomly allocated to one of the eight groups. Participants were tasked with assembling a LEGO car in 71 steps, spread out over three different workstations (B in Fig. 1). 
The visual support received depended on the group participants were allocated to. 
For all groups, a large virtual monitor in the VE displayed the assembly instructions. 
These were developed based on, and closely resembling, LEGO’s official instruction manual (A in Fig. 1). 
Figure 1 shows the concrete implementation within the VE. 
Lastly, upon completing the assembly task, participants progressed to the post-questionnaire.

\subsection{Measurements}
Time to completion (TTC) and mistakes were metered automatically in the VE. Both the attempted placement of the correct part at an incorrect location or an incorrect part at any location was counted as mistake. Cognitive load (CL) was measured using NASA-TLX~\cite{Hart1988, yoghourdjian_2021_scalability}.
Attention check questions (ACQs) and mutually exclusive questions were also included to ensure participants' attentiveness and the coherence of their answers. 
The VE was developed using Unity based on an adapted version used in previous studies and exported as a webVR application~\cite{Farr2023, Bohne2022, zuercher2022optimising}.

\section{Results}
A total of 199 participants were included in this study (for raw data see \cite{rawdata}). 63\% of participants were male and the average age was 34.8 years (SD = 9.8 years). As their occupation, 74 participants selected 'full-time MTurk worker', 72 selected 'office worker', 29 selected 'non-office worker', and 9 selected 'university student'. As their highest completed educational background, 52 indicated 'high school', 134 'college', and 13 'other'. An added analysis of the participants’ experience with LEGO and VR revealed a homogeneous distribution across all experience levels.

Based on the collected data, we statistically analysed the dependent variables.
To determine the suitable test, we tested for normal distribution using the Shapiro-Wilk test \cite{shapiro1965analysis} and homogeneity of variance \cite{brown1974robust}.
The assumption of normality was invalidated by two out of three distributions. For normal distributions we applied ANOVA, and for non-normal distributions, we applied the Kruskal-Wallis (KW) test \cite{kruskal1952use} and calculated the average values ($\mu$) using the median \cite{kruskal1952use, brown1974robust}.

\subsection{Time to Completion}

\begin{figure}[h]
    \centering
    \includegraphics[width=0.45\textwidth]{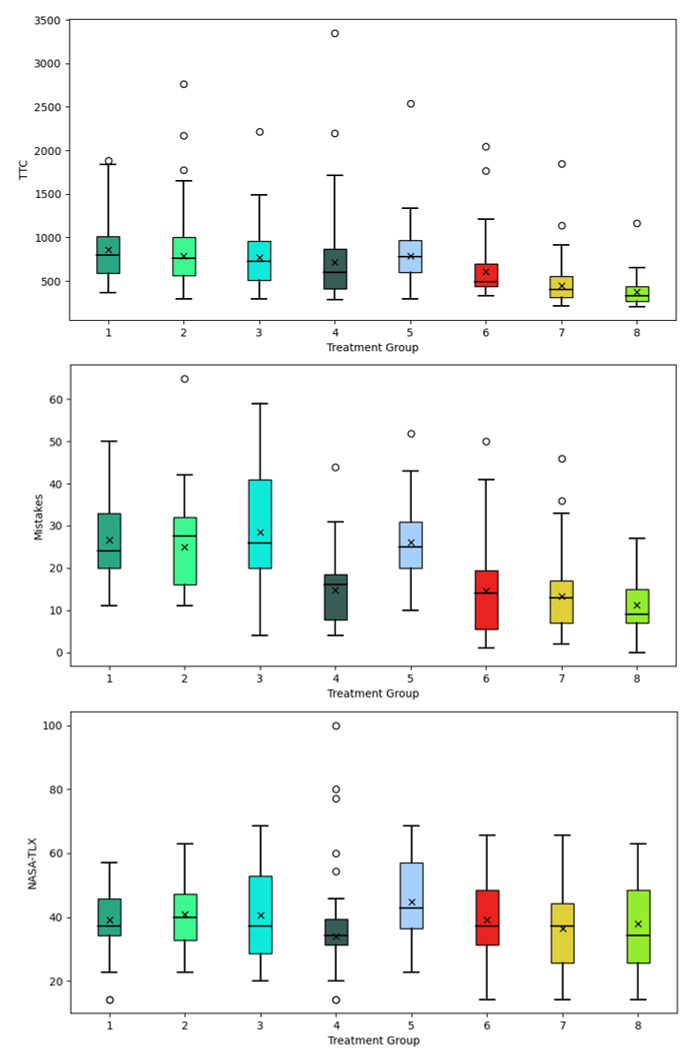}
    \caption{Distribution plots for task completion times, mistakes made and cognitive load by group (from left to right)}
    \label{fig:my_label}
\end{figure}

The first hypothesis (H1) was confirmed, as our results and the non-parametric KW test ($H=62.153, p<0.001, f^2=0.449$) demonstrate that the average TTC differs significantly between the visualisation types (see \autoref{fig:my_label}).
The descriptive statistics indicate that the control group had the highest TTC (G1; $\mu$=797s), followed by
combined GG and OI (G5; $\mu$=778s),
GG (G2; $\mu$=759s),
OI (G3; $\mu$=723s),
AG (G4; $\mu$=597s),
combined GG and AG (G6; $\mu$=493s),
combined OI and AG (G7; $\mu$=403s), and
combined GG, OI, and AG (G8; $\mu$=333s).

\subsection{Number of Mistakes Made}

\noindent The second hypothesis (H2) was confirmed, as the ANOVA test shows a significant difference in the number of mistakes made between the visualisation types ($F=12.827, p<0.001, f^2=0.506$).
Our descriptive statistics reveal that combined GG, OI, and AG (G8; $\mu$=9) lead to the least number of errors followed by
combined OI and AG (G7; $\mu$=13),
combined GG and AG (G6; $\mu$=14),
AG (G4, $\mu$=16),
the control group (G1; $\mu$=24),
combined GG and OI (G5; $\mu$=25),
OI (G3;$\mu$=26), and
GG (G2; $\mu$=27.5). 

\subsection{Cognitive Load}

The third hypothesis (H3) was not confirmed, as the KW test indicates no significant differences in cognitive load between the groups ($H=8.28, p=0.309, f^2=0.007$) on the basis of the NASA-TLX data.
The lowest cognitive load was found in combined GG, OI, and AG (G8; $\mu$=34) and AG (G4; $\mu$=34), followed by combined OI and AG (G7; $\mu$=37), OI (G3; $\mu$=37), combined GG and AG (G6; $\mu$=37), and the control group (G1; $\mu$=37). The highest load was found in GG  (G2; $\mu$=40) and combined GG and OI (G5; $\mu$=43).

 \subsection{Qualitative Feedback}

Most participants who provided free-text answers were satisfied with the implemented XR Visual Guidance and confirmed our initial assumptions of their helpfulness to them during the different stages of the assembly process: \emph{“Easy to understand with arrows pointing to objects and destinations highlighted”}. Multiple participants from those groups lacking one of the three dimensions of Visual Guidance pointed out a need for the lacking dimension – without knowing about the existence of the other groups. This, therefore, supports the overall design of the XR Visual Guidance framework by Pietschmann et al. \cite{Pietschmann2022}. Participants without GG, for example, reported problems with orientation, such as locating the next workstation: \emph{“I also had a little trouble locating the new station, perhaps [an] arrow showing where the new station was would have been helpful.”} Participants without OI reported problems identifying the correct next piece, especially when there were multiple pieces which looked alike: \emph{“Highlighting would have been nice, especially in cases where there were multiple identical components [...]”}. Participants without AG suggested that situated visualisation, supporting them in finding the correct placement location, would have been beneficial: \emph{“It would be helpful to provide translucent installation instructions in the field of view.”} However, there were also complains about the design of the OI arrow, as it was not precise enough and sometimes obstructed the field of view: \emph{“I think highlighting the piece that needs moving would work better than having an arrow pointing to it.”} Additionally, a handful of participants complained about performance issues, a lack of clear instructions, and cyber sickness.
Overall, the experiment and especially the implemented XR Visual Guidance were well-received. 

\section{Discussion}

The aim of this large-scale, immersive study was to quantify the impact of different forms of XR Visual Guidance on user performance by investigating time to completion and mistakes made, as well as the underlying human factors, such as cognitive load.

\subsection{XR Visual Guidance Positively Impacting User Performance}

H1 and H2 were confirmed, as the results of our study affirmed a significant difference between the individual visualisations for both TTC and the number of mistakes made. Visual Guidance was able to reduce TTC by as much as 58\% and mistakes by 62\% when comparing G8 against the control group (G1). The overall differences between the groups for TTC and mistakes are strongly significant (p<0.001). Interestingly, TTC was reduced most by combining OI and AG, whereas the reduction in mistakes appeared to be largely associated with AG only. GG did not appear to have a strong overall impact, which is not surprising given that tridimensional orientation was only a minor part of the assembly task. 

Based on the above observations, AG, which is designed to support the placement process, appears to have effectively mitigated challenges associated with the placement process, thus leading to a shorter and more accurate placement of parts. AG alone (G4) was able to reduce TTC by 25\% and mistakes by 33\%. 
OI, which is designed to help the user pick the correct part, by itself (G3) seems to have shortened the time to identify and select the correct part by 9\%. Surprisingly, OI did not appear to have a big impact on mistakes made, thus indicating that users did not have any issues with selecting the wrong parts.
Combining both AG and OI (G7) yields a reduction in TTC and mistakes made by up to 49\% and 46\% respectively. In line with our expectations, providing the users with visuals to help them navigate the most challenging parts of the assembly process, such as choosing the right part and placing it where it belongs, yielded the biggest improvements in performance.

The above findings on TTC contrast recent findings from~\cite{Jasche2021} who reported that the type of visualisation did not impact TTC. 
However, they align with previous studies presenting significantly reduced task completion times using different forms of visualisation~\cite{Blattgerste2018, Smith2020}.
Regarding mistakes, the above findings are in line with~\cite{Jasche2021, Blattgerste2018, Smith2020}.

\subsection{Cognitive Load not Impacted by Different XR Visual Guidance}
H3 was not confirmed since there is no significant difference in cognitive load between the individual visualisations. Although CL was reduced in G7 compared to the control group (G1), overall the results did not cross the significance threshold (p=0.27). 

The similarity of cognitive load in combination with stark differences in user performance comes as a surprise, as the increase in performance does not appear to be the result of a reduction in cognitive load based on our observations. We expected to see an increase in performance with reduced cognitive load associated with a reduction of the split-attention effect. However, this does not seem to be the case. Here, further research is needed in this regard, as this finding contrasts with those of previous studies~\cite{Jasche2021, Blattgerste2018}.

\subsection{Practical Takeaways}

Our study yielded multiple takeaways for leveraging and designing XR Visual Guidance.
\textbf{Firstly, XR Visual Guidance can significantly increase user performance if designed and applied systematically.} 
Instead of copying textual information from standard operating procedures and displaying them without adaptation, those designing immersive applications, especially in industrial contexts, should use the unique opportunity of embedding visualisations offered by XR technology. 
Pietschmann et al's underlying XR Visual Guidance framework can further guide such design processes~\cite{Pietschmann2022}.
\textbf{Secondly, Action Guidance most efficiently reduces mistakes.} 
In combination with Object Identification, a strong reduction in TTC is also possible. For practitioners, this means that identifying the type of challenge their users are facing is instrumental in selecting an appropriate Visual Guidance technique. For example, the placement process seemed to be a source of mistakes for our participants, and AG was able to mitigate associated challenges.
\textbf{Finally, Object Identification can reduce picking times but should not obstruct users' fields of view.} 
While the overall effect on performance was positive, the most frequent constructive criticism participants provided was that parts to be selected should be highlighted without obstructing their field of view. Based on this, highlighting approaches utilising glow or colour change techniques appear to be more advisable than arrows pointing out the correct objects. Furthermore, this emphasises the potential drawbacks of occlusion and visual clutter.

\subsection{Limitations and Opportunities for Future Research}

While the amount and diversity of our recruited participants were one of the greatest strengths of this study, controlling the experimental environment and ensuring the validity of the collected data remained a challenge due to the remote setup. Although we were able to check the plausibility of the data, participants nonetheless took part in the experiment in an uncontrolled environment on their own machines. One upside of this circumstance, however, is that this reflects a more realistic setup than those recreated in lab-based immersive experiments; thus, our research may potentially yield insights that are more reflective of real-world engagements with applied XR Visual Guidance. Although we had several technical checks in place, we could not entirely rule out poor machine performance or distractions during the experiment. 
Furthermore, measurements of human factors in this study are subjectively queried. While questionnaires such as NASA-TLX, IMI, and SUS are highly standardised, future work should aim to additionally include biometric data collected via different sensors, such as eye tracking or EEG technology.

Among potential avenues for further research, two themes stand out in particular: Firstly, the relationship between performance improvements and the underlying human factors remains to be investigated. In this regard, we were not able to obtain results in line with our expectations, with previous studies reporting mixed evidence as well. Especially the causes and effects of cognitive load in combination with XR Visual Guidance need to be better understood to design efficient visual support and leverage the unique opportunities offered by XR technology to augment human workforces.

Secondly, exploring additional design options of Gaze Guidance, Object Identification, and Action Guidance, or even the entire XR Visual Guidance framework, could lead to further insights into effective XR Visual Guidance. While we chose promising approaches based on well-performing design options from previous studies, the variety of visualisations we could implement as part of this study was limited. Higher variation in Visual Guidance methods used may ultimately lead to more holistic design guidelines for future XR designers and developers. Further investigating the impact different visualisations may have on performance and human factors would likely lead to novel and interesting insights.

\section{Conclusion}

The relevance of well-designed Visual Guidance and its importance for enhancing human performance is increasingly being recognised within a growing XR community. While previous studies have investigated the impact of different forms of Visual Guidance on user performance, using exploratory small-scale experiments, we set out to quantify the impact on user performance based on a larger-scale study and a highly diverse participant pool. Based on Pietschmann et al’s XR Visual Guidance framework, we designed eight different groups with differing forms of XR Visual Guidance \cite{Pietschmann2022}. Our results confirmed that XR Visual Guidance can reduce the time to complete a virtual assembly task by as much as 58\% and the mistakes made by 62\%. While user performance greatly benefited from the implemented XR Visual Guidance, cognitive load did not seem to differ. Our data suggested that improvements in performance were not linked to a decrease in cognitive load; this is a significant insight, given the deviating results of related studies. Our findings demonstrate the potential of XR Visual Guidance in visually immersive environments to improve user performance, particularly for tasks requiring procedural knowledge. This study paves the way for a deeper exploration of the underlying human factors associated with XR Visual Guidance and the relationship between visualisation, performance, and cognitive load.


\acknowledgments{The authors wish to thank K. E. Ruf for the invaluable suggestions and support during the development of this publication, as well as K. Ciezarek for his work on the virtual environment. Furthermore, we wish to thank the RADMA Association and Hanns Seidel Foundation for their continued support and research sponsoring.
This research is supported in part by the Engineering and Physical Sciences Research Council’s AgriFoRwArdS program [EP/S023917/1] as well as NSF award III-2107328 and NIH award R01HD104969. }

\bibliographystyle{abbrv-doi}
\bibliography{ref}

\end{document}